\documentclass{article}
\usepackage{graphicx}
\usepackage{epsfig}
\usepackage{amsfonts}
\usepackage{amsmath, amssymb, array}
\usepackage{graphics,graphicx,float,mfpic}
\usepackage{epsfig}
\usepackage{color}
\newcommand{\be}{\begin{equation}}
\newcommand{\ee}{\end{equation}}
\newcommand{\bea}{\begin{eqnarray}}
\newcommand{\eea}{\end{eqnarray}}
\newcommand{\bmat}{\begin{pmatrix}}
\newcommand{\emat}{\end{pmatrix}}

\title{Non Uniform Black Strings and Critical Dimensions in $AdS_d$}
\author{{\large T\'erence Delsate\footnote{terence.delsate@umh.ac.be} }\\ \\{\small Theoretical and Mathematical Physics Dept.}\\{\small University of Mons-Hainaut, Belgium}}
\date{\today}

\begin{document}
\maketitle

\begin{abstract}
We study the equations of black strings in spacetimes of arbitrary dimensions with a negative cosmological constant and construct numerically non uniform black strings solutions. Our results suggest the existence of a localised black hole in asymptotically locally $AdS$ spacetime. 
We also present evidences for a dependence of the critical dimension on the horizon radius.The critical dimension represents the dimension where the order of the phase transition between uniform and non uniform black string changes. Finally, we argue that both, the regular asymptotically locally $AdS$ solution and $AdS$ black string solutions with a very small horizon radius, present a negative tension. This turns out to be an unexpected feature of the solutions.
\end{abstract}

\section{Introduction}
The last decade has witnessed a growing interest for solutions in $AdS$ spacetime. Part of the reason is the $AdS/CFT$ \cite{adscft, adscft2} duality in string theory. This duality relates classical $AdS$ gravity solutions to operators in a conformal field theory defined in the background of the boundary $AdS$ spacetime. Among others, an application of $AdS/CFT$ duality allowed to interpret the transition between the thermal $AdS$ spacetime and the $AdS$ black hole as a transition from a confining to a deconfining phase in the dual CFT \cite{hptr}. 
At some regime in the dual $CFT$, the theory is approximated by a relativistic fluid dynamic theory. Many works have been recently carried out in order to investigate different fluids configurations. This hydrodynamical approach provides information on the expected properties of the solutions on the gravity side (see for example \cite{hydrocft, hydrocft2}). 
It is clear that $AdS$ spacetime are interesting from the $AdS/CFT$ point of view, but they are also interesting by theirselves, since unexpected properties can arise, due to the $AdS$ curvature effect \cite{pnubsd}. The duality is better understood in five dimensions, where the $AdS_5$ space is dual to a $\mathcal N=4$ Super Yang-Mills theory. Clearly, $AdS/CFT$ and even more, string theory involves higher dimensional models for gravity.
But it turned out that in more than four spacetime dimensions and without a cosmological constant, various theorems known to hold in four dimensions are not true anymore. For instance, the uniqueness theorem doesn't hold anymore in higher dimensional gravity, where different black hole solutions for a given set of physical parameters can exist as it is the case for the Myers-Perry black hole and the black rings \cite{solubr1}. In four dimension, the uniqueness theorem imposes only $S_2$ horizon topology, while in higher dimensions, apart from the black hole solutions with horizon topology $S_{d-2}$ \cite{solubh, solubh1, solubh2, solubh3, solubh4, solubh5, solubh6}, one can construct black strings solutions with horizon topology $S_{d-3}\times S_1$ \cite{solubs1, solubs2, solubs3, rms} but also solutions with toroidal horizon topology \cite{solubr1,solubr2}.

However, the black string solutions are known to suffer from a long wavelength instability, namely the Gregory-Laflamme instability \cite{gl}. This instability is somehow the counterpart of the Plateau-Rayleigh instability in fluid dynamics. After that discovery in '93, it was widely believed that the black string would decay to a localised black hole, but it has been argued that such a phase transition would take an infinite proper time at the horizon \cite{horowitz}. This lead to postulate the existence of an intermediate solution, the non uniform black string, which has been constructed first in a perturbation theory \cite{gubser} then in the full non linear regim \cite{wiseman}.

More recently, it has been shown that the Gregory-Laflamme instability persists in asymptotically locally $AdS$ spacetimes \cite{rbd}, and $AdS$ non uniform black strings have been constructed perturbatively in various dimensions \cite{pnubsads, pnubsd}. In \cite{pnubsd}, partial results were presented for the non perturbative non uniform black string solution in $AdS_6$.

In this paper, we investigate the non perturbative non uniform black string solutions in various dimensions. The paper is organised as follows: first we present the model, the equations and boundary conditions for the non uniform black string problem in section \ref{model}. In section \ref{properties}, we give the asymptotic and near horizon solution for the non uniform black strings, as well as the thermodynamical properties of the solutions. Section \ref{properties} also includes a discussion on the sign of the tension for small $AdS$ black strings, as well as a discussion on the entropy comparison between uniform solutions and non uniform solutions with the same mass. We describe the numerical method and results in section \ref{num}. It is well known that in asymptotically locally flat spacetime the order of the phase transition between the uniform and non uniform phase changes from one to higher above a critical dimension \cite{sorkin}, but  we will argue that this property changes drastically in asymptotically locally $AdS$ spacetime. All our results are summarised in the concluding section \ref{conclusion}.

\section{The Model}
\label{model}
We consider the $d$-dimensional Einstein-Hilbert action with the presence of a negative cosmological constant $\Lambda$, supplemented by the Hawking Gibbons Boundary term and with :
\be
S = \frac{1}{16\pi}\int_{\mathcal M} \sqrt{-g}\left( R+\frac{(d-1)(d-2)}{\ell^2} \right)d^dx - \frac{1}{8\pi}\int_{\partial\mathcal M} \sqrt{-h}Kd^{d-1}x
\label{ehact}
\ee
in units where the Newton constant equals one, $G_d=1$ and where $\mathcal M$ is the manifold under consideration, $\partial\mathcal M$ is the boundary of the manifold, $g$ is the determinent of the metric, $R$ is the scalar curvature, $\ell$ is the $AdS$ radius related to the cosmological constant according to  $\Lambda = -(d-1)(d-2)/(2\ell^2)$, $h$ is the determinent of the induced metric on the boundary and $K$ is the trace of the extrinsic curvature of the boundary manifold.

In order to construct the non uniform strings, we supplement the action \eqref{ehact} by the following ansatz
\be
ds^2 = -b(r)e^{2A(r,z)}dt^2 + e^{2B(r,z)}\left(\frac{dr^2}{f(r)} + a(r)dz^2\right) + r^2e^{2C(r,z)}d\Omega_{d-3}^2
\label{ansnubs}
\ee
where $a,b,f$ refer to the background functions, solution to the uniform black string problem (where $A=B=C=0$)  \cite{rms}  and $A,B,C$ are the non uniform corrections to the background. The functions $a,b,f$ behave like $r^2/\ell^2$ for large $r$:
\be
a(r)\approx b(r)\approx f(r)\approx \frac{r^2}{\ell^2} + \mathcal O(1),
\label{asbg}
\ee
describing an asymptotically locally $AdS$ spacetime.

However, the solutions are a priori independent of the $a,b$ functions normalisation; this means that the functions $a,b$ are to be multiplied by a suitable factor in order to follow the asymptotic behaviour \eqref{asbg} {\it a posteriori}; this invariance is in fact the freedom one has to rescale the time coordinate or the $z$ coordinate. In order to avoid the {\it a posteriori} normalisation, we perform the following change of functions
\be
a(r) = \frac{r^2}{\ell^2}\hat a(r),\ b(r) = \frac{r^2}{\ell^2}\hat b(r),\ f(r) = \frac{r^2}{\ell^2}\hat f(r).
\label{chfun}
\ee
The advantage of this change of functions is that the hatted functions normalisation is fixed once for all, independently of the number of dimension and of the $AdS$ radius by imposing $\hat a\approx \hat b \rightarrow 1$ as $r\rightarrow\infty$. This normalisation doesn't require an \emph{a posteriori} rescaling of the solution.

Note that the $z$ coordinate ranges from $0$ to some length $L$. The value of $L$ will be fixed to be the critical length where the linear pertubations are static. This precise value of $L$ is provided by the stability analysis coming from the first order in perturbation theory \cite{rbd}.

It must be stressed that the equations resulting from the ansatz \eqref{ansnubs}  require nontrivial regularity conditions for the correction functions equations. However, going in a 'conformal-like' gauge,
\be
ds^2 = \frac{g(\tilde r)}{\ell^2}\tilde b(\tilde r)e^{2A(r,z)} dt^2 + e^{2B(r,z)}\left(\frac{\ell^2dr^2}{g(\tilde r)^2 \tilde f(\tilde r)}+\frac{g(\tilde r) L^2\tilde a(\tilde r)dz^2}{\ell^2}  \right) + g(\tilde r)e^{2C(r,z)}d\Omega_{d-3}^2,
\label{confgauge}
\ee
where $g(r)=\tilde r^2 + r_h$, $r_h$ being the horizon radius, leads to much simpler regularity conditions. Note that we rescaled the $z$ coordinate such that $z\in[0,1]$ by factorising the length $L$.

The relation between \eqref{ansnubs} and \eqref{confgauge} is given by
\be
\tilde r^2 + r_h^2 = r^2,\ \tilde a(\tilde r) = a(r),\ \tilde b(\tilde r) = b(r),\ \tilde f(\tilde r) = f(r).
\ee


\subsection{The equations and Boundary Conditions}
From now on, the $a,b,f$ functions and $r$ coordinate will refer to the functions and variables of the conformal gauge \eqref{confgauge}. The Einstein equations with the ansätz \eqref{confgauge} lead to the following set of coupled partial differential equations:
\bea
&\left.\right.& -\left(\frac{\left(-1+d\right)e^{2B}r^2}{f{g}^2}\right)-\frac{b'}{2rb}+\frac{a'b'}{4ab}-\frac{{b'}^2}{4{b}^2}+\frac{b'f'}{4bf}-
\frac{g'}{2rg}+\frac{a'g'}{4ag}+\frac{\left(2+d\right)b'g'}{4bg}+\frac{f'g'}{4fg}\nonumber\\
&\left.\right.&+\frac{\left(-1+d\right){g'}^2}{4{g}^2}+\frac{b''}{2b}+\frac{g''}{2g}+\frac{{\ell}^4r^2{A^{(0,1)}}^2}{L^2af{g}^3}+ \frac{\left(-3+d\right){\ell}^4r^2A^{(0,1)}C^{(0,1)}}{L^2af{g}^3}+\frac{{\ell}^4r^2A^{(0,2)}}{L^2af{g}^3}-\frac{A^{(1,0)}}{r}\nonumber\\ &\left.\right.&+\frac{a'A^{(1,0)}}{2a}+\frac{b'A^{(1,0)}}{b}+\frac{f'A^{(1,0)}}{2f}+\frac{\left(2+d\right)g'A^{(1,0)}}{2g}+{A^{(1,0)}}^2+\frac{\left(-3+d\right)b'C^{(1,0)}}{2b}\nonumber\\
&\left.\right.&+\frac{\left(-3+d\right)g'C^{(1,0)}}{2g}+\left(-3+d\right)A^{(1,0)}C^{(1,0)}+A^{(2,0)}=0
\label{eqA}
\eea

\bea
&\left.\right.&\frac{\left(-4+d\right)\left(-3+d\right)e^{2B-2C}{\ell}^2r^2}{2f{g}^3}+\frac{\left(-4+d\right)\left(-1+d\right)e^{2B}r^2}{2f{g}^2}-\frac{a'}{2ra}-\frac{{a'}^2}{4{a}^2}+\frac{a'f'}{4af}-\frac{g'}{2rg}+\frac{a'g'}{ag}\nonumber\\
&\left.\right.&-\frac{\left(-3+d\right)b'g'}{4bg}+\frac{f'g'}{4fg}-\frac{\left(-4+d\right)\left(-1+d\right){g'}^2}{8{g}^2}+\frac{a''}{2a}+\frac{g''}{2g}- \frac{\left(-3+d\right){\ell}^4r^2A^{(0,1)}C^{(0,1)}}{L^2af{g}^3}\nonumber\\
&\left.\right.&-\frac{\left(-4+d\right)\left(-3+d\right){\ell}^4r^2{C^{(0,1)}}^2}{2L^2af{g}^3}+\frac{{\ell}^4r^2B^{(0,2)}}{L^2af{g}^3}-\frac{\left(-3+d\right) g'A^{(1,0)}}{2g}-\frac{B^{(1,0)}}{r}+\frac{a'B^{(1,0)}}{2a}+\frac{f'B^{(1,0)}}{2f}\nonumber\\
&\left.\right.&+\frac{3g'B^{(1,0)}}{2g}-\frac{\left(-3+d\right)b'C^{(1,0)}}{2b}-\frac{{\left(-3+d\right)}^2g'C^{(1,0)}}{2g}-\left(-3+d\right)A^{(1,0)}C^{(1,0)}-\frac{\left(-4+d\right)\left(-3+d\right){C^{(1,0)}}^2}{2}\nonumber\\
&\left.\right.&+B^{(2,0)}=0
\label{eqB}
\eea

\bea
&\left.\right.&-\left(\frac{\left(-4+d\right)e^{2B-2C}{\ell}^2r^2}{f{g}^3}\right)-\frac{\left(-1+d\right)e^{2B}r^2}{f{g}^2}-\frac{g'}{2rg}+\frac{a'g'}{4ag}+\frac{b'g'}{4bg}+\frac{f'g'}{4fg}+\frac{\left(-1+d\right){g'}^2}{4{g}^2}\nonumber\\
&\left.\right.&+\frac{g''}{2g}+\frac{{\ell}^4r^2A^{(0,1)}C^{(0,1)}}{L^2af{g}^3}+\frac{\left(-3+d\right){\ell}^4r^2{C^{(0,1)}}^2}{L^2af{g}^3} +\frac{{\ell}^4r^2C^{(0,2)}}{L^2af{g}^3}+\frac{g'A^{(1,0)}}{2g}-\frac{C^{(1,0)}}{r}+\frac{a'C^{(1,0)}}{2a}\nonumber\\
&\left.\right.&+\frac{b'C^{(1,0)}}{2b}+\frac{f'C^{(1,0)}}{2f} +\frac{\left(-1+d\right)g'C^{(1,0)}}{g} +A^{(1,0)}C^{(1,0)} +\left(-3+d\right){C^{(1,0)}}^2+C^{(2,0)}=0
\label{eqC}
\eea
where the exponent $(m,n)$ denotes the $m^{th}$ derivative with repect to $r$ and $n^{th}$ derivative with respect to $z$ and the primes denote the derivative with respect to $r$ for functions depending on $r$ only.

The system of partial differential equations above is supplemented by the following boundary conditions:
\bea
A(r,z)&=&B(r,z)=C(r,z)=0\mbox{ for } r\rightarrow\infty,\nonumber\\
\partial_z A(r,z)&=&\partial_z B(r,z)=\partial_z C(r,z)=0\mbox{ for }z=0,1,\\
\partial_r A(r,z)&=&\partial_rC(r,z)=0,\ B(r,z)=A(r,z) + d_0\mbox{ for } r=0\nonumber,
\eea
where $d_0$ is a real deformation parameter related to the variation of the temperature along the non uniform branch. The first condition ensures the fact that $A,B,C$ are \emph{corrections} and thus vanish at infinity, the second conditions imposes periodicity along the $z$ coordinate while the last three conditions are the regularity conditions.


\section{Properties of the solution}
\label{properties}
\subsection{Asymptotic and near horizon behaviour}

The asymptotic behaviour of the background functions is given in \cite{rms}. Since we factorise the $\frac{r^2}{\ell^2}$ term, the asymptotic developpement in \cite{rms} is of course to be divided by $r^2/\ell^2$, leading to

\bea
a&=& 1  + \sum_{i=0}^{\left\lfloor \frac{d-4}{2}\right\rfloor} a_i\left(\frac{\ell}{r}\right)^{2i+2} + c_z\left(\frac{\ell}{r}\right)^{d-1} + \sum_k\delta_{d,2k+1}\xi\log\frac{r}{\ell}\left(\frac{\ell}{r}\right)^{d-1} + \mathcal O\left(\frac{\ell}{r}\right)^{d},\nonumber\\
b&=& 1  + \sum_{i=0}^{\left\lfloor \frac{d-4}{2}\right\rfloor} a_i\left(\frac{\ell}{r}\right)^{2i+2} + c_t\left(\frac{\ell}{r}\right)^{d-1} + \sum_k\delta_{d,2k+1}\xi\log\frac{r}{\ell}\left(\frac{\ell}{r}\right)^{d-1} + \mathcal O\left(\frac{\ell}{r}\right)^{d},\\
f&=& 1  + \sum_{i=0}^{\left\lfloor \frac{d-4}{2}\right\rfloor} f_i\left(\frac{\ell}{r}\right)^{2i+2} + (c_t+c_z+c_0)\left(\frac{\ell}{r}\right)^{d-1} + \sum_k\delta_{d,2k+1}\xi\log\frac{r}{\ell}\left(\frac{\ell}{r}\right)^{d-1} + \mathcal O\left(\frac{\ell}{r}\right)^{d},\nonumber
\label{asbckg}
\eea
where $\delta$ is the Kronecker delta, $\lfloor X \rfloor$ stands for the floor integer value of $X$, $c_t,c_z$ are parameters to be determined numerically and $a_i,f_i,c_0,\xi$ depend on $d$ and are given in \cite{rms}.

It has been argued in \cite{pnubsads,pnubsd} that the modes of $A,B,C$ in a perturbation theory decay as $\left(\frac{\ell}{r}\right)^{d-1}$ in all order of the perturbation theory. Since the non perturbative solution is a combination of these modes, the leading order should also decay as
\be
A(r,z)=A_1\left(\frac{\ell}{r}\right)^{d-1} + \mathcal O\left(\frac{\ell}{r}\right)^{d},\ B(r,z)=B_1\left(\frac{\ell}{r}\right)^{d-1} + \mathcal O\left(\frac{\ell}{r}\right)^{d},\ C(r,z)=C_1\left(\frac{\ell}{r}\right)^{d-1} + \mathcal O\left(\frac{\ell}{r}\right)^{d}.
\label{decaynu}
\ee
for some $A_1,B_1,C_1\in\mathbb R$. The decay \eqref{decaynu} is confirmed by direct computations.

Close to the horizon, where $r\approx r_h$, the background fields obey the following series developement to the leading order
\bea
a(r)&=& a_0 + \mathcal O\left(r-r_h\right),\ b(r)= b_1 (r-r_h) + \mathcal O\left(r-r_h\right)^2,\nonumber\\
f(r)&=& \frac{1}{r_h^3}\left((d-1)r_h^2 + (d-4)\ell^2\right) (r-r_h) + \mathcal O\left(r-r_h\right)^2,
\eea
where $a_0,b_1$ are real numbers fixed by the background boundary condition $a,b,f\rightarrow1$ for $r\rightarrow\infty$.
Subleading corrections can be found in \cite{rms}, where the coefficients are to be multiplied by $\ell^2/r_h^2$ due to the change of functions \eqref{chfun}.


\subsection{Thermodynamical properties}
The temperature, mass, entropy and tension of the background is given by \cite{rms}:
\bea
T_U &=&\frac{1}{4\pi}\sqrt{\frac{b_1}{r_h\ell^2}\left((d-1)r_h^2  + (d-4)\ell^2\right)} ,\ S_U=\frac{r_h^{d-1}}{4\ell^2}a_0LV_{d-3} ,\nonumber\\ 
M_U &=&\frac{L\ell^{d-4}V_{d-3}}{16\pi}\left(c_z - (d-2)c_t\right) + M_U^0  ,\ \mathcal T_U= \frac{\ell^{d-4}V_{d-3}}{16\pi}\left((d-2)c_z - c_t\right)+\mathcal T_U^0,
\eea
where $M_U^0 = \mathcal T_U^0 L$ are Casimir-like terms occuring in odd dimensions \cite{rms}.

The thermodynamical properties of the background has been studied in \cite{rms} and the thermodynamical properties of the non uniform strings has been studied perturbatively in \cite{pnubsads,pnubsd}. The correction functions provide corrections on the thermodynamical quantities computed from the background (as the name correction function suggests). The quantities computed on the background solution are denoted with a subscript $U$, while the quantities refering to the entire non uniform solution will be denoted with a subscript $NU$. The corrections on the thermodynamical quantities are then given by
\bea
T_{NU} &=& e^{-d_0} T_U\nonumber,\\
S_{NU} &=& \frac{S_U}{L}\int_0^1 L e^{2(B(r_h,z)+(d-3)C(r_h,z))}dz,
\eea
where $T_{NU}$ and $S_{NU}$ are the Hawking temperature (resp. the entropy) of the non uniform string. Note that the first relation makes the role of $d_0$ clearer: once the background is set, the parameter $d_0$ allows to slide along the non uniform branch in a temperature-entropy phase diagram.

These quantities are defined at the horizon and are relatively easy to compute from a technical point of view. The asymptotic quantities are given by
\bea
M_{NU} &=& M_U\left(1+\frac{\delta M}{M_U} \right) ,\nonumber\\
\mathcal T_{NU} &=& \mathcal T_U\left( 1+\frac{\delta \mathcal T}{T_U} \right).
\eea
where $\delta M, \delta\mathcal T$ are functions of the parameters $A_1,B_1,C_1$ appearing in \eqref{decaynu} and $c_t,c_z,c_0$.
Note that at the boundary spacetime, $\partial_z$ and $\partial_t$ are both killing vectors, so the definition of the mass and tension as functions of the asymtptotic parameters \cite{rms} remain the same. Note also that the decay \eqref{decaynu} combined with the asymptotic background solution gives contribution of order $r^{-(d-3)}$ to $g_{tt}, g_{zz}$ and leads to correction on $c_t,c_z$, thus on the mass and tension; another decay of the corrections would then be physically rejectable.

However, the quantities $c_t,c_z$ and worse, $A_1,B_1,C_1$ are difficult to compute, first because of the sharpness of the decay and second because the determination of these quantities is plagued by numerical noise. In order to avoid this problem, we were able to integrate the first law of black holes in order to compute the masses (of the uniform solution \emph{and} of the non uniform solution). Assuming that the Smarr relation $M + L\mathcal T = T_H S$, holds for the non uniform solution, we were able to compute the corrections on the tension for the non uniform solution.

The procedure to integrate the first law is performed in two steps, first at the background level, then for the non uniform solution. 
At the background level, given the mass per unit length for a solution, say with $\ell=1, r_h=r_h^*$ and given the entropy and temperature as a function of $r_h$, we can integrate the first law of black objects according to
\be
M_U(r_h)/L = \int_{r_h^*}^{r_h} T_H(r_h)\frac{\partial S_U(r_h)}{\partial r_h} dr_h/L,
\ee
where we have assumed $L$ to be a constant. This assumption doesn't spoil generality since $L$ plays a spectator role in the background.

The tension is computed using the Smarr relation once the mass is known:
\be
\mathcal T_U = \frac{M_U}{L} - T_H^U\frac{S_U}{L}.
\ee

The integration of the fist law for the non uniform solution is more straightforward since we already compute \emph{corrections} as a function of the parameter $d_0$. The procedure to integrate the first law seems more natural in this case; given a uniform black string with horizon radius $r_h$, the non uniform black string with a given value of $d_0$ emanating from this uniform black string has a mass given by
\be
M_{NU}(r_h) = M_U(r_h) + \int_0^{d_0} T_H^{NU}\frac{\delta S}{\delta d_0}d d_0 = M_U(r_h) + T_H^U\int_0^{d_0} e^{-d_0}\frac{\delta S}{\delta d_0}d d_0 ,
\ee
since the length of the non uniform black string is fixed and where $\delta S/\delta d_0$ is of course to be evaluated numerically.

Once the mass of the non uniform string for the given value of $d_0$ computed, it is again straightforward to obtain the tension for the corresponding value of $d_0$, assuming the Smarr relation:
\be
\mathcal T_{NU} = \frac{T_H^{NU}S_{NU}}{L} - \frac{M_{NU}}{L}.
\ee


\subsubsection{Note on the tension of the background}

\begin{figure}[H]
   \center 
    \includegraphics[scale=.6]{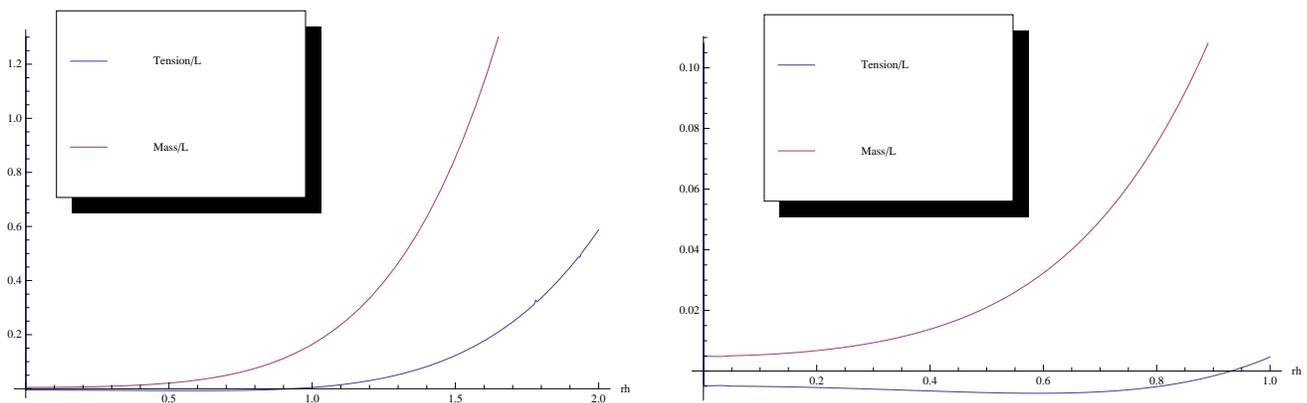}
    \caption{The mass and tension per unit length for the uniform black string in $d=6, \ell=1$. Large values of $r_h$ exhibit a positive tension while smaller values of $r_h$ have negative tension.}
    \label{fig:d6mt}
  \end{figure}

\begin{figure}[H]
   \center 
    \includegraphics[scale=.6]{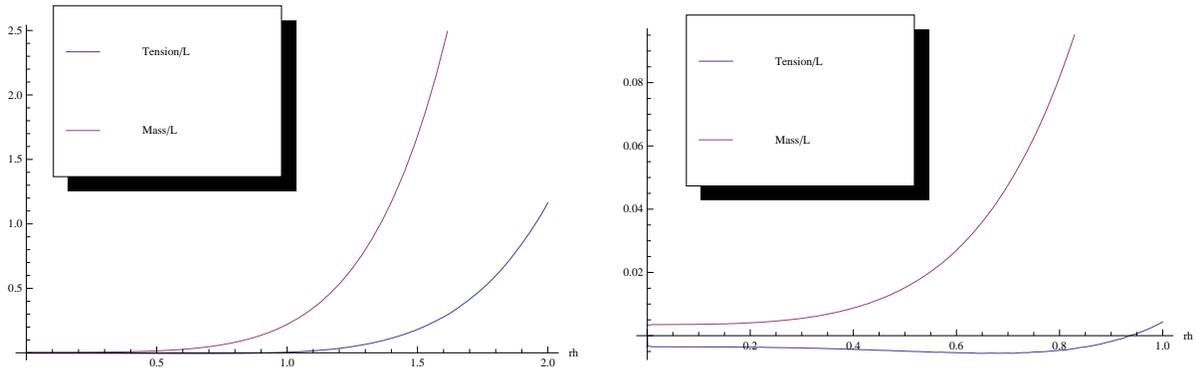}
    \caption{This figure is the same as \ref{fig:d6mt}, except that here, $d=7$.}
    \label{fig:d7mt}
  \end{figure}

A regular solution in the asymptotically locally flat spacetime would be the $d-1$ dimensional Minkowski spacetime times a circle, $\mathcal M_{d-1}\times S^1$. Of course, this solution has zero mass and zero tension. The $d$ dimensional uniform black string in the flat case is characterised by a positive mass and tension. However, the $AdS$ counterpart of the regular solution is characterised by a positive non vanishing mass \cite{rms}, a finite temperature and a vanishing entropy. Thus, using the smarr relation, the tension of the regular solution is negative:
\be
M + \mathcal T L = 0\ \Rightarrow \mathcal T = - \frac{M}{L}.
\ee
It turns out that the black strings with small values of $r_h$ also have negative tension, these black strings are continuously connected to the $AdS$ regular solution. This is confirmed by a numerical analysis as shown in figure \ref{fig:d6mt} and \ref{fig:d7mt} for $d=6, \ell=1$ and $d=7, \ell=1$. These values were obtained using the definition of the mass and tension as a function of $c_t$, $c_z$ in \cite{rms} (in units where the Newton constant is set to one). Note however that the regular solution ($r_h=0$) has a vanishing total energy since $E = M + \mathcal T L = 0$, where $E$ denotes the total energy.

The negative tension seems to be an effect of the $AdS$ curvature, since this phenomenon doesn't appear in the asymptotically locally flat uniform string problem. This might be interpreted as follows: a negative cosmological constant acts like antigravity, and since the tension is essentially a binding energy, the small uniform black strings (small values of $r_h$) are not massive enough to admit a positive binding energy, due to the balance of the $AdS$ repulsive effect.

We checked the sign of the tension numercially for $d=6,7$ but we expect this property to hold for every value of the number of dimension.

\subsubsection{Entropy Difference at same mass}
In the asymptotically locally flat spacetime case, it is possible to evaluate the entropy difference at same mass between the non uniform and uniform string using the following relation (see for example \cite{sorkin, gubser}):
\be
\frac{S_{\mbox{Non uniform}}}{S_{\mbox{Uniform}}} = 1 + \sigma_1 \lambda^2 + \sigma_2 \lambda^4 + ...,
\ee
where $\sigma_1, \sigma_2$ are functions of the corrections on the temperature, entropy, mass and tensions computed in a perturbative approach and $\lambda$ is the deformation parameter in this perturbative approach \cite{sorkin}.

This relation permitted to establish the existence of a critical dimension above which the order of the phase transition between the uniform black string and the non uniform one changes from one to higher.

However, in order to establish this kind  of relation, one uses the mass dependence of the entropopy of the $(d-1)$ dimensional Tangherlini black hole, since the latter are foliations at fixed $z$ of the uniform black string (in asymptotically locally flat spacetime). However in asymptotically locally $AdS$, one cannot rely on the same argument since the solution for the uniform black string is numerical and the relation between the mass and entropy of the constant $z$ foliation is not clear in this case; so one has to rely on numerical computations.

At fixed value of the $AdS$ length, from a numerical point of view, the mass and entropy in the uniform phase are functions of $r_h$, $M(r_h)$ (resp. $S(r_h)$) while the corrections on the mass and on the entropy along the non uniform string are functions of $r_h$ and $d_0$, say $\delta M(r_h,d_0)$ (resp. $\delta S(r_h,d_0)$). So one can compute the variation of the mass along the non uniform branch and find the value of $r_h$, say $r_h^*$, such that the uniform string has the same mass. Once this value is found, it is trivial to compute the difference between the entropy of the non uniform phase emerging from $r_h$ and with a given value of $d_0$ and the entropy of the uniform string with horizon radius $r_h^*$:
\be
r_h^* \mbox{is such that } M(r_h) + \delta M(r_h,d_0) = M(r_h^*).
\ee
Then, the relative entropy difference between the non uniform string with horizon radius $r_h$ and the uniform string of same mass is a function of $d_0$ and is given by
\be
\frac{S_{\mbox{Non uniform}}-S_{\mbox{Uniform}}}{S_{\mbox{Uniform}}} = \frac{(S(r_h) + \delta S(r_h,d_0))-S(r_h^*)}{S(r_h^*)}.
\label{dsnuu}
\ee

We will use the relation \eqref{dsnuu} in order to investigate the order of the phase transition. If the entropy of the uniform string is higher than the entropy of the non uniform one for the same value of the mass in a small deformation regime, the order of the phase transition is $1$. If the entropy of the non uniform string increases at the same mass, the order is higher than one and the transition is smooth. In other words if the quantity evaluated from \eqref{dsnuu} is negative, the order of the phase transition in one, if the quantity is positive, the order is higher.

We will present the result of this investigation in the case where $d=7$ in the next section.


\subsection{Relation between deformation and thermodynamics}
\label{defthlink}
The order parameter for the transition from the uniform string phase to the non uniform string is defined by \cite{gubser}
\be
\lambda = \frac{R_{Max}}{R_{Min}}-1,
\ee
where $R_{Min}$ (resp. $R_{Max}$) is the maximum (resp. minimum) of the areal radius, defined as $R = r e^{C(r,z)}$.

The parameter $\lambda$ measures the deformation of the solution; the uniform string is caracterised by $\lambda = 0$, while the hypothetic localised black hole phase has an infinite deformation parameter. It has been argued in \cite{flatphase} that using the mass and tension is more relevant in the construction of the phase diagram; the mass and tension remain finite in the three phases (uniform, non uniform and localised black hole/string). However, the mass and tension are defined at the boundary spacetime and are difficult to compute from a technical point of view. An alternative is to study the phase diagram in $T_H,\ S$ coordinates; these quantities are defined at the horizon of the black objects, and are easier to compute.

As already noted in \cite{pnubsd}, $\lambda$ increases with $d_0$ when the cosmological constant increases. This can be understood as follows: the deformation is related to the variation of entropy and $d_0$ is related to the variation of the temperature. Moreover the rate of variation of $\lambda$ with $d_0$ is proportional to the varation of the entropy with respect to the temperature along the non uniform phase. In order to prove this explanation, one should recall that for small values of $\lambda$, the non uniform string is well described by the first order of perturbation theory (where $C(r,z)\approx \epsilon C_1(r)\cos 2\pi z$, $\epsilon$ being the small parameter of the perturbation). In this approximation, one can fix the value of $\epsilon$ for a given value of $\lambda$:
\be
\lambda = \frac{e^{\epsilon C_1(r_h)} + \mathcal O(\epsilon)^2}{e^{-\epsilon C_1(r_h))+ \mathcal O(\epsilon)^2}} -1= 2\epsilon C_1(r_h) + \mathcal O(\epsilon)^2.
\ee
Recall that from the boundary conditions of the first order in perturbation theory \cite{rbd}, $C_1(r_h)=1$, leading to $\epsilon = \lambda/2$.

On the other hand, the variation of the temperature along the non uniform branch is given by $T_H^{NU} = T_U e^{-d_0}$; a variation $\delta d_0$ in $d_0$ gives rise to a variation of the temperature $\delta T_H^{NU}$ in the non uniform phase according to $\delta T_H^{NU} = -T_H^Ue^{-d_0}\delta d_0$. Moreover, the correction on the entropy in the perturbative approach appears in order $\epsilon^2$ in the perturbation theory; in other words, is proportional to $\lambda^2$. 

So, increasing $d_0$ from $0$ to $d_0<<1$ implies a variation of $\lambda$ of $0$ to $\lambda<<1$ and
\be
\frac{\delta S}{\delta T_H^{NU}} = -T_H^{NU}\frac{\lambda^2}{d_0} \approx T_H^U\frac{\lambda^2}{d_0}.
\ee
In other words, the value of $\lambda$ corresponding to a (small) value of $d_0$ is a monotonic negative function of  $\frac{\delta S}{\delta T_H^{NU}}$.

\begin{figure}[H]
   \center 
    \includegraphics[scale=.4]{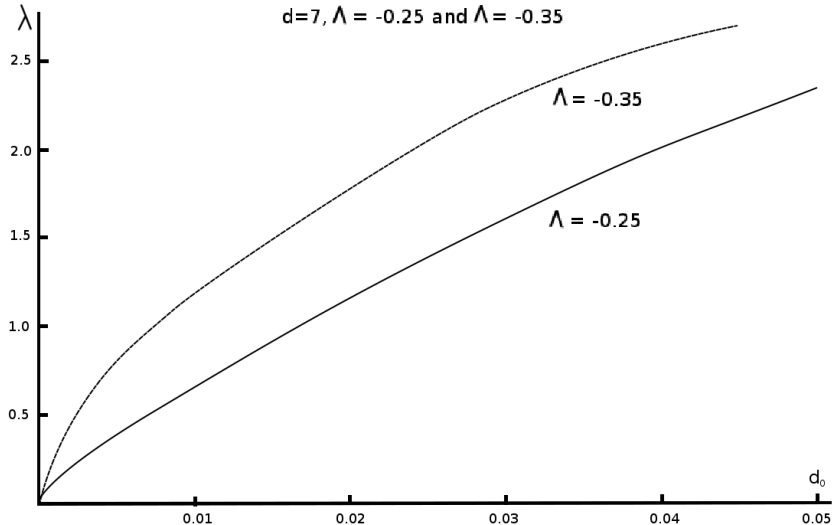}
    \caption{The deformation increases faster for larger absolute values of the cosmological constant, this is to be related to the increase of $\delta S/\delta T$ with $\Lambda$ (large absolute  
     values of $\Lambda$ at fixed $r_h$ correspond to large values of $r_h$ at fixed $\Lambda$ from the scaling relations in \cite{rms}).}
    \label{fig:d7def}
  \end{figure}

To sum up, if $\frac{\delta S}{\delta T_H^{NU}}$ is large and negative, $\lambda$ will be large and positive and conversely. This was noticed but not explained in \cite{pnubsd} and is shown for $d=7$  for two values of $\Lambda$ in figure \ref{fig:d7def}.

\section{Numerical Results and Discussion}
\label{num}
\subsection{Numerical Procedure}
In order to integrate the equations \eqref{eqA}, \eqref{eqB}, \eqref{eqC} we first integrate the background equations and the first order of perturation theory numerically in order to set the background and to fix the length of the static perturbation, where the non uniform phase should emerge. The integration is carried out numerically with the solver \emph{Colsys} \cite{COLSYS}. Then, we integrate the system of partial differential equation numerically, using the solver \emph{FidiSol} \cite{fidi} with the suitable boundary conditions described in the previous section.
Let us mention the fact that the numerical procedure is plagued by many difficulties. First, the background functions are numerical and it is not clear how the errors on the background influe errors on the corrections.

Moreover, one cannot start from the trivial solution $A=B=C=0$; Fidisol needs an initial guess sufficently close to the solution. In practise, it is sufficient to use the solution to the first order in pertubation theory as a starting profile. 

Let us also mention that if the parameters are not choosen in a suitable manner, the solver produces a uniform solution which seems to be a 'correction' to the uniform string leading to a uniform string with a different temperature. Of course, we are not interested in this kind of solutions. Moreover, it happened that the solution is a kind of combination of this uniform correction nd some non uniform correction, leading to a decrease of the deformation along the non uniform string, even for small values of the cosmological constant (or equivalently for small values of the horizon radius at fixed $AdS$ radius) where we expect the pattern of the solution to be similar to the case where the cosmological constant vanishes. 

The region of large horizon radius is even more difficult to investigate : we expect a transition from thermodynamically unstable non uniform black strings to thermodynamically stable non uniform black strings (refered to as small-short and small-long non uniform black strings in \cite{pnubsd}). In the boundary separating these two phases, the small-short and small-log black string is characterised by $\delta S/\delta T_H \rightarrow\infty$. As we argued in section \ref{defthlink} the deformation would become extremely high even for very small values of $d_0$ (thus of $\delta T_H$). Of course, this extreme behaviour is not manageable from a technical point of view and sets the limit of our numerical investigation.

However, we were able to find some hints for the existence of non uniform solutions in a neighbourhood of this region, but these solutions have large numerical errors and it is not reliable to perform quantitative computations.

\subsection{Numerical solutions}
For the reason described above, we restrict the study of the $AdS$ non uniform black string to the region of the parameter space where $r_h<<\ell$. We also need a criterium to decide wether we accept or reject solutions. The first criterium is of course that the solution is a deformed solution, we are interested in non uniform corrections. Second, we expect the deformation to increase, so we reject solutions with decreasing deformation (when $d_0$ increases). However, it must be stressed that the numerical error increases with the deformation: in the limit where the string is extremely deformed, the slope of the corrections becomes very high and leads to numerical unprecisions.

\begin{figure}[H]
   \center 
    \includegraphics[scale=.3]{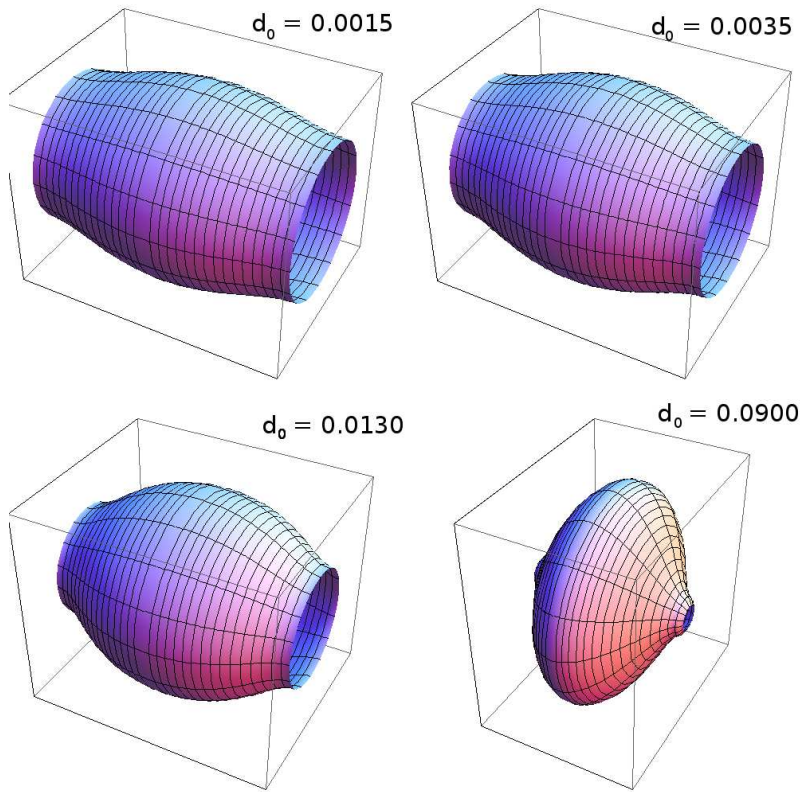}
    \caption{Embedding of the horizon of non uniform black strings with $d=7$, $\Lambda=-0.25$ in a 3 dimensional euclidean space. The horizon looks like a black hole horizon as the deformation increases.}
    \label{fig:d7hor}
  \end{figure}

Despite of all the numerical difficulties, we were able to construct several solutions for various values of $d$, $\Lambda$ and $d_0$. In every case considered, the deformation is an increasing function of $d_0$ and we systematically found extremely deformed solution, providing strong indications for the existence of the $AdS$ localised black hole phase. In figure \ref{fig:d7hor}, we present embeddings of the event horizon in a 3 dimensional cartesian space; the strings is almost uniform for small values of $d_0$ and looks like a sphere touching the borders for larger values of $d_0$.

\subsection{Horizon radius depending critical dimension}
\begin{figure}[H]
   \center 
    \includegraphics[scale=.5]{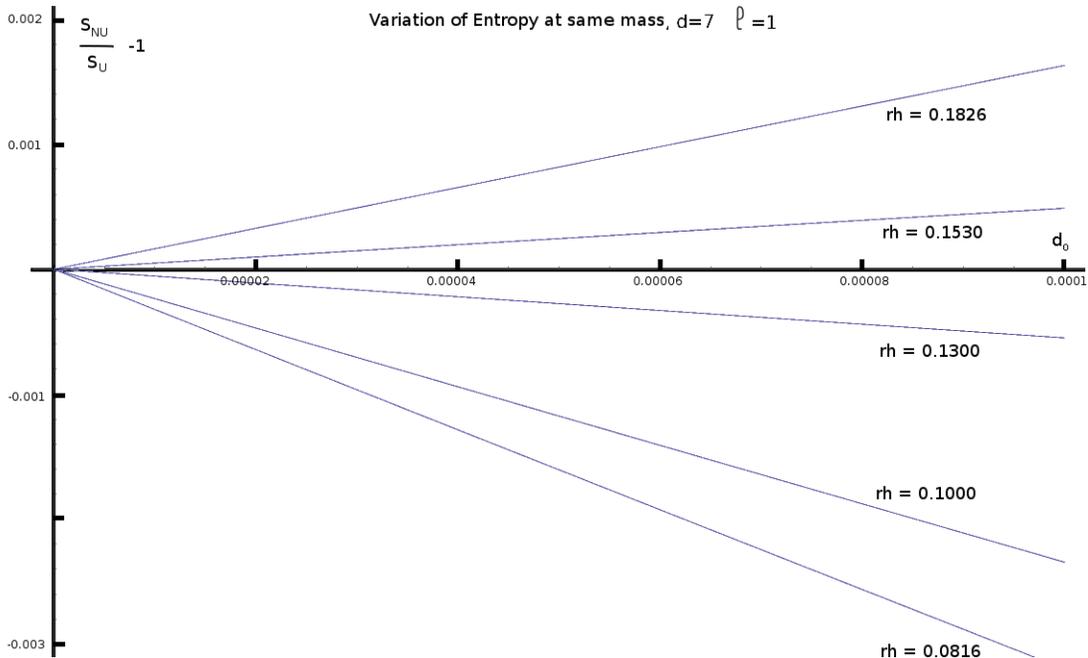}
    \caption{The relative entropy difference between the non uniform phase and uniform phase at same mass as a function of $d_0$ for $d=7,\ell=1$ and various value of $r_h$.}
    \label{fig:criticalrh}
  \end{figure}

We have evaluated the entropy difference between the non uniform black string and the uniforms black string at same mass and found that above some critical values of the horizon radius, the uniform black string is more entropic than the non uniform one while for larger valures of $r_h$, the non uniform black string becomes more entropic. In other words, the phase transition can be of first order or of higher order for the same value of the dimension, depending on the value of the horizon radius. We present a plot of the variation of the entropy difference as a function of $d_0$ for $d=7, \ell=1$ in figure \ref{fig:criticalrh} for various values of the horizon radius. Small values have a negative slope while larger value acquire a positive slope, changing the order of the phase transition.

We expect this property to hold for all values of the dimension.


\section{Conclusion}
\label{conclusion}

We have constructed the non uniform asymptotically locally $AdS$ black string for various values of the number of dimensions (typically $5,6,7,8$). We were able to numerically integrate the system of coupled partial differential equations despite of the numerical difficulties and found numerical evidences for the existence of the non uniform black string. Moreover, the variation of the deformation of the solutions with respect to the temperature is closely related to the variation of the entropy with temperature in the small deformation region. This of course makes the investigation of the small-long $AdS$ non uniform black string \cite{pnubsd} very difficult from a technical point of view. However, the small short $AdS$ black strings are easier to deal with and we found strong evidence for the existence of a $AdS$ localised black hole phase in the strongly deformed limit of the non uniform black string for the dimensions we considered. From a qualitative point of view, the solutions behave similarly in all the dimensions we considered, so we expect all these properties to hold for an arbitrary number of dimensions.

Moreover, it might have been expected from the fact that the variation of the entropy with respect to the variation of the temperature increases with $r_h$, that the critical dimension for the order of the phase transition of the uniform - non uniform black string depends on $r_h$, at fixed $AdS$ radius. It seems indeed that this is the case. We numerically compared the entropy of non uniform black strings and uniform strings with same mass for $d=7$ and found that above some value of $r_h$ ($r_h\approx 0.14$ for $\ell=1$) the entropy along the non uniform branch increases quicker than the entropy of the uniform string with same mass. In other words, the transition becomes smooth, thus of order higher than one. Once again, we expect this property to hold for other numbers of dimensions, at least for $d<d^*$ where $d^*$ is the critical dimension reported in \cite{sorkin}.

We also argued that the asymptotically locally $AdS$ regular solution has a negative tension and a positive mass. It was already noticed that the masss was positive in \cite{rms}, but we stress the fact that the tension is negative. The tension is also negative for uniform black strings with small values of the horizon radius (compared to the $AdS$ radius). However, the total energy for the regular solution is null and positive for the small $AdS$ black strings. We interpret this property as resulting from a balance between the attractive effect of the mass and the repulsive effect of the $AdS$; if the mass is too small, the repulsive effect dominates and the string is not bound (the tension is interpreted as a binding energy when multiplied by the length of the $z$ coordinate).

Note that the various new properties we exposed here should influence the dual $CFT$ in the $AdS/CFT$ context.

\section{Acknowledgement}
I would like to gratefully acknowledge Yves Brihaye and Eugen Radu for very useful discussions and wise advices. I also thank Fabien Buisseret and Georges Kohnen.

\bibliography{biblio}{}
\bibliographystyle{unsrt}
\end{document}